\documentclass[pra,twocolumn,floatfix,notitlepage,superscriptaddress,longbibliography]{revtex4-1}
\usepackage{amsmath}

\usepackage[T1]{fontenc}
\usepackage[utf8]{inputenc}
\usepackage{amsfonts}
\usepackage{bbm}
\usepackage{bm}
\usepackage{graphicx}
\usepackage[bookmarks,bookmarksopen,bookmarksdepth=2]{hyperref}
\hypersetup{
    colorlinks=true,
    citecolor=black,
    linkcolor=black,
    filecolor=black,
    urlcolor=black,
}
\usepackage[capitalise]{cleveref}
\usepackage{mathtools}
\usepackage{mathrsfs}
\usepackage{physics}
\usepackage{etoolbox}
\usepackage{datetime2}
\usepackage{dsfont}
\usepackage{tikz}
\usetikzlibrary{decorations.pathreplacing,calligraphy,backgrounds}
\usepackage[utf8]{inputenc}

\newcommand{\nocontentsline}[3]{}
\newcommand{\tocless}[2]{\bgroup\let\addcontentsline=\nocontentsline#1{#2}\egroup}

\newcommand{\SU}{\mathrm{SU}}
\makeatletter
\DeclareRobustCommand{\element}[1]{\@element#1\@nil}
\def\@element#1#2\@nil{%
  #1%
  \if\relax#2\relax\else\MakeLowercase{#2}\fi}
\makeatother

\begin{document}
\widetext

\title{Multispin Clifford codes for angular momentum errors in spin systems}

\author{Sivaprasad Omanakuttan}
\email[]{somanakuttan@unm.edu}
\affiliation{Center for Quantum Information and Control (CQuIC), Department of Physics and Astronomy, University of New Mexico, Albuquerque, New Mexico 87131, USA}
\author{Jonathan A. Gross }
\email[]{jarthurgross@google.com}
\affiliation{Google Quantum AI, Venice, CA 90291, USA
}

\date{\today}
\begin{abstract}

The physical symmetries of a system play a central role in quantum error correction.
In this work we encode a qubit in a collection of systems with angular-momentum symmetry (spins), extending the tools developed in \cite{gross_designing_2021} for single large spins.
By considering large spins present in atomic systems and focusing on their collective symmetric subspace, we develop new codes with octahedral symmetry capable of correcting errors up to second order in angular-momentum operators.
These errors include the most physically relevant noise sources such as microwave control errors and  optical pumping.
We additionally explore new qubit codes that exhibit distance scaling commensurate with the surface code while permitting transversal single-qubit Clifford operations.


\end{abstract}
\maketitle
\section{Introduction\label{sec:intro}}
Quantum error correction (QEC) is an essential ingredient for implementing  quantum computation reliably. 
In simple words, QEC uses a large Hilbert space to encode a smaller-dimensional system to overcome the detrimental effects of decoherence and recover the ideal state of an encoded system.
One standard strategy for QEC, analogous to classical error correction, where the major error is the bit flip, is to encode a qubit of information in multiple qubits.
However, due to the fact that for QEC one needs to account for both bit flip and phase flip errors, the number of physical qubits required to encode a logical qubit is very large.
In spite of this difficulty, these techniques are widely considered for QEC and have found a lot of success including recent experimental implementation using the surface codes and color codes~\cite{acharya2022suppressing,ryan2022implementing,krinner2022realizing}.

Another approach for QEC is to encode a qubit in a single system with a large Hilbert space; for example, the standard GKP code where a qubit is encoded in a simple harmonic oscillator, whose large Hilbert space provides natural protection from many errors native to this system~\cite{PhysRevA.64.012310,PhysRevA.97.032346}. 
This approach in general reduces the overhead and thus makes the scaling easier. 
There have been many recent ideas about quantum computation using GKP states~\cite{Bourassa2021,xu2022qubit,noh2020encoding,omanakuttan2022spin} and a recent experiment where real-time quantum error correction beyond break-even is demonstrated~\cite{sivak2022real}.

In~\cite{gross_designing_2021}, quantum error-correcting codes native to spin systems with spin larger than $1/2$ were developed using the special symmetries associated with these systems. 
In particular, the binary octahedral symmetry was used; however, one needs a very large spin ($j\geq13/2$) to build a fully error-correcting code for this symmetry.
In this work, we find a way out of this need for big spins by using the tensor product of multiple spins for spin larger than $j=1/2$ and using the irreducible $\SU(2)$ representations in the symmetric subspace of these tensor products. 
These systems could generally be of great potential as they are easier to scale and systems with an order of $100$ spins have been used for quantum simulation experiments with neutral atoms~\cite{ebadi2021quantum,scholl2021quantum}.
In spin systems, the main source of decoherence is random rotations which contribute to the first-order errors in angular momentum and optical pumping which is a second-order effect in angular momentum  involving vector and tensor light shifts~\cite{deutsch2010quantum,omanakuttan2021quantum}. 
Accordingly, designing codes in these composite spin systems that correct for first- and second-order angular-momentum errors could reduce the overhead required to achieve fault-tolerant regimes of quantum computation and thus accelerate the path to useful quantum computation.

Similarly, we also consider the case of the tensor product of qubit systems.
We encode a qubit in the symmetric subspace of multiple qubits to find codes that have transversal Cliffords and correct arbitrarily large errors. 
Using the binary octahedral symmetry we demonstrate explicit codewords with distance $3$ and distance $5$, and generally find that the minimum number of qubits required for a given distance scales similarly to the surface code while allowing full single-qubit transversal Clifford operations.

The remainder of this article is organized as follows.
In \cref{sec:introduction_to_binary_octahedral_code} we gave a brief introduction to the binary octahedral code and the natural symmetry associated with these quantum error-correcting codes.
In \cref{sec:Knill-Laflamme_condition} we study the Knill-Laflamme condition for a general spin system by using the spherical tensor operators.
In \cref{sec:su_2_irreps} we find the relevant $\SU(2)$ irreps in the symmetric subspace for the tensor product of spin systems by mapping it to bosons.
We used these approaches to find useful codes that correct for first-order angular momentum   (small random $\mathrm{SU}(2)$) errors in \cref{sec:correcting_first_order_errors} and the second-order (Light shift) errors in \cref{sec:correcting_second_order_errors}.
In \cref{sec:symmetric_subspace_1_2}, we study how one can apply these approaches to the tensor product of multiple spins $j=1/2$ (qubit) systems and create error-correcting codes in the symmetric subspace of this multipartite system, finding explicit codes with distance $3$ and $5$.
We give the outlook and possible future directions in \cref{sec:conlusions_and_future_work}.

\section{Introduction to binary octahedral code}
\label{sec:introduction_to_binary_octahedral_code}

We build upon work~\cite{gross_designing_2021} done to encode information against random $\SU(2)$ rotations in large single spins (irreps of $\SU(2)$).
This task is simplified by restricting ourselves to codespaces that are preserved under the action of a finite subgroup of $\SU(2)$, such as the single-qubit Clifford group (binary octahedral group).
If the finite subgroup is rich enough, the full set of Knill-Laflamme conditions for first-order rotation errors reduces to a single expectation value, which is simple to check.
The single-qubit Clifford group is one such rich subgroup, in that it can map any of $\{J_x, J_y, J_z\}$ to any other, with either sign.
These symmetries allow one to consolidate the conditions to
\begin{align}
    \bra{i}_{}J_z\ket{j}_{}
    &=
    C_{0z}\delta_{ij}
    \\
    \bra{i}_{}J_xJ_y\ket{j}_{}
    &=
    C_{xy}\delta_{ij}
    \\
    \bra{i}_{}J_z^2\ket{j}_{}
    &=
    C_{zz}\delta_{ij}
    \,.
\end{align}
The fact that a $\pi$ rotation about $J_z$ must put a relative phase between logical 0 and 1 means that one must have ``odd'' support on the $J_z$ basis states and the other must-have ``even'' support, which further reduces the conditions to
\begin{align}
    \bra{0}J_z\ket{0}
    &=
    0
    \,.
\end{align}

It turns out the binary tetrahedral group (a subgroup of the binary octahedral group) has enough symmetries for the above argument to go through as well, so we will also consider codes with that symmetry in this work.

The binary octahedral group, having additionally the $S$ gate, a $\pi/2$ rotation about $J_z$, further constrains the support of the codewords in the $J_z$ basis, such that the $J_z$ eigenvalues included in logical 0 are either $4\mathbf{Z}+\tfrac{1}{2}$ or $4\mathbf{Z}-\tfrac{3}{2}$, depending on the code, and the eigenvalues for logical 1 are the negatives.

\section{Derivation of Knill-Laflamme conditions}
\label{sec:Knill-Laflamme_condition}
\begin{figure}
    \centering
     \includegraphics[width=0.25\textwidth]{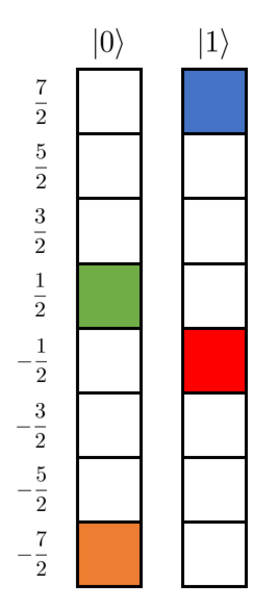}
    \caption{The codewords $\ket{0}$ and $\ket{1}$ for the $\varrho_4$ irrep of the binary ocathedral symmetry for $j=7/2$ in the angular momentum basis.
    The colored boxes indicate the states occupied whereas the blank ones indicate those states are not occupied for the codeword. 
    The states in the codeword are spaced by four units of angular momentum $m_z=\pm 4$, a standard property of the octahedral symmetry, and contribute to the error correction condition.
    The codewords $\ket{0}$ and $\ket{1}$ are separated a single unit of angular momentum and hence overlap of $\bra{0}T^{k}_1\ket{1}=(-1)^k\bra{1}T^{k}_{-1}\ket{0} \neq 0 $  for odd values of $k$ whereas $\bra{0}T^{k}_{-1}\ket{1}=(-1)^k\bra{1}T^{k}_{1}\ket{0} = 0 $. 
    This contributes to the off-diagonal terms to consider for error correction in \cref{eq:knill_Laflamme_conditions}.
    }
    \label{fig:code_figure}
\end{figure}
In this section, we extend the Knill-Laflamme condition derived for small random $\SU(2)$ rotations in large single spins in \cite{gross_designing_2021} to general errors which are powers of angular momentum operators.
Since products of angular-momentum operators up to a given order are not linearly independent (due to equivalence relations such as the commutation relations), it can be convenient to use spherical tensors~\cite{sakurai2014modern,klimov2008generalized,chinni2022reliability} as an error basis:
\begin{multline}
T^{k}_q(j)=
\\
\sqrt{\frac{2k+1}{2j+1}}\sum_m \braket{j,m+q}{k,q;j,m} \ketbra{j,m+q}{j,m}
\end{multline}
which are basically the sums of powers of the angular momentum operators and are related to spherical harmonics.
Using this as our basis of errors, the Knill-Laflamme conditions \cite{PhysRevA.55.900} require that
\begin{align}
    \bra{i}E_a^\dagger E_b\ket{j}
    &=
    \delta_{ij}C_{ab}
    \\
    E_a,E_b
    &\in
    \{T^{k}_q\}_{0\leq k\leq N}
\end{align}
if we want to be able to correct angular-momentum errors of orders up to $N$.
Because products of spherical tensors are sums of spherical tensors
\begin{align}
    T^{k}_qT^{k^\prime}_{q^\prime}
    &=
    \sqrt{(2k+1)(2k^\prime+1)}\sum_{\tilde{k}} c^{\tilde{k}}_{\tilde{q}}T^{\tilde{k}}_{\tilde{q}}
\end{align}
where $\tilde{q}=q+q^\prime$ and that the sum over $\tilde{k}$ is restricted over $|k-k^\prime|\leq \tilde{k} \leq k+ k^\prime$ and $c^{\tilde{k}}_{\tilde{q}}$ is defined in terms of $6j$ symbols and Clebsch-Gordon coefficients~\cite{klimov2008generalized}
\begin{equation}
    c^{\tilde{k}}_{\tilde{q}}=(-1)^{2j+\tilde{k}}  \left\{ \begin{array}{c c c}
    k & k^\prime & \tilde{k} \\
    j & j & j \end{array}\right\} C_{k,q ,k^{\prime} q^\prime}^{\tilde{k} \tilde{q}} .
\end{equation}
We can equivalently consider the conditions
\begin{align}
    \bra{j}T^{\tilde{k}}_{\tilde{q}}\ket{k}
    &=
    \delta_{jk}C^{\tilde{k}}_{\tilde{q}}
    \\
    0
    &\leq
    \tilde{k}
    \leq
    2N
    \,.
\end{align}

Consider the unitary $U_{X}=\exp(-i\pi J_x)$ the octahedral symmetry of the  states gives us, an overall global phase that is irrelevant,
\begin{equation}
\begin{aligned}
U_{X}\ket{0}=&\ket{1}\\
U_{X}\ket{1}=&\ket{0}
\end{aligned}
\label{eq:Ux_property}
\end{equation}
and we can find that 
\begin{equation}
    U_{X} T_{q}^{k} U_{X}^{\dagger}=(-1)^{k}T^{k}_{-q}
\end{equation}
where the details of this calculation are given in \cref{sec:spherical_tensors}.
Using this we see that for the codewords
\begin{equation}
\bra{0}T_q^{k}\ket{0}=(-1)^k\bra{1}T_{-q}^{k}\ket{1}
\end{equation}
For the case of the code words with octahedral symmetry, the code words are real in the angular-momentum basis (see \cref{sec:real_code_word}) and so is $T_q^{k}$ and thus when we have two states $\ket{\psi}$ and $\ket{\phi}$ which are real linear combinations of the code words that respect the binary octahedral symmetry,
\begin{equation}
\bra{\psi}T^{k}_{-q}\ket{\phi}=(-1)^q\bra{\phi}T^{k}_{q}\ket{\psi}
    \label{eq:property_2}
\end{equation}
which we prove in \cref{sec:real_code_word}.
Thus one gets,
\begin{equation}
\begin{aligned}
\bra{0}T_q^{k}\ket{0}=(-1)^k\bra{1}T_{-q}^{k}\ket{1}=(-1)^{k-q}\bra{1}T_{q}^{k}\ket{1}
\end{aligned}
\end{equation}
so from the above equation, the error condition is trivially satisfied unless
\begin{equation}
\left(k-q\right) \text{ mod 2}=1
\end{equation}
However, the code words have support on the $J_z$ eigenstates that are separated by $q \text{ mod 4}=0$, as described in \cref{sec:introduction_to_binary_octahedral_code} and is given in \cref{fig:code_figure}, and hence the expression is identically zero unless $q$ is even and thus the only diagonal conditions we need to check are those when $k$ is odd:
\begin{equation}
\{T_{0}^{1},T_{0}^{3},T^{5}_{0},T^{5}_{4},\hdots\}
\end{equation}

Now thinking about the next error-correction condition we get
\begin{equation}
\begin{aligned}
\bra{0}T_q^{k}\ket{1}=&(-1)^{k}\bra{1}T_{-q}^{k}\ket{0}\\
=& (-1)^{k-q}\bra{0}T_q^{k}\ket{1}
\end{aligned}
\end{equation}
The above equation states that when 
\begin{equation}
k-q \text{ mod }2=1
\end{equation}
we automatically get 
\begin{equation}
\bra{0}T_q^{k}\ket{1}=0
\end{equation}
Now again the support of the different code words is separated by odd shifts in angular momentum and hence we also automatically get that
\begin{equation}
\bra{0}T_q^{k}\ket{1}=0
\end{equation}
when $q$ mod $4=1$ and can be seen from \cref{fig:code_figure}.
Thus the only off-diagonal conditions we need to check are when both $k$ and $q$ are odd
\begin{equation}
\{T^{1}_{1},T_{1}^{3},T_{-3}^{3},T^{5}_{5},T^{5}_{1},T^{5}_{-3},\hdots\}
\end{equation}
Hence the error correction conditions can be written as,
\begin{widetext}
\begin{equation}
\begin{aligned}
\bra{0}T_q^{(k)}\ket{0}=&(-1)^{(k-q)} \bra{1}T_q^{(k)}\ket{1}\implies \text{ only consider } (k \in \text{ odd and }q \equiv 0\text { mod } 4),\\
\bra{0}T_q^{(k)}\ket{1}=&(-1)^{(k-q)} \bra{0}T_q^{(k)}\ket{1}\implies \text{ only consider } (k \in \text{ odd and }q\equiv 1 \text { mod }4).
\end{aligned}
\label{eq:knill_Laflamme_conditions}
\end{equation}
\end{widetext}

This gives the general error correction conditions one needs to check for the binary octahedral codes. 
One can easily see that a large number of conditions are trivially satisfied accounting for the symmetry of the codewords. 
In the following sections, we will see how these correction conditions will help us in obtaining useful quantum-error-correction codes.

\section{The $\SU(2)$ irreps in the symmetric subspace of the tensor product of $n$ spin $j$ systems}
\label{sec:su_2_irreps}



Now, consider the tensor product of $n$ spin $j$ systems.
This forms a Hilbert space $\mathcal{H}$ of dimension $d^n$ where $d=2j+1$. 
We focus on the symmetric subspace \cite{harrow2013church} where expectation values are unchanged by permuting the subsystems, so for any arbitrary operators $A_1, A_2,\dots, A_n$ we have
\begin{multline}
     \langle A_1\otimes A_2\otimes \cdots \otimes A_n\rangle
     =
     \\
     \langle A_{\pi(1)}\otimes A_{\pi(2)} \otimes \cdots \otimes A_{\pi(n)} \rangle.
\label{eq:permuatation_symmetry}
\end{multline}
for any permutation $\pi$.
Restricting our attention to the symmetric subspace simplifies the Knill-Laflamme conditions, as many of the error terms $E_a^\dagger E_b$ that arise are permutations of each other and need only be verified once within the symmetric subspace.

The dimension of the symmetric subspace for the tensor product of $n$ spin-$j$  systems is,
\begin{equation}
\mathrm{dim}\left(S_n(d)\right)=\frac{d(d+1)...(d+n-1)}{n!}.
\label{eq:dim_sym_space}
\end{equation}

Since we are interested in encoding qubits in the symmetric subspace, we need to identify how the symmetric subspace decomposes into $\SU(2)$ irreps.
For $j=1/2$ the decomposition is simple, as the symmetric subspace is itself a spin-$(n+1)/2$ irrep.
For larger spins, we must work harder, as the symmetric subspace decomposes into multiple $\SU(2)$ irreps.

One way to see that we must get multiple $\SU(2)$ irreps in the symmetric subspace is to notice that the operator $J_z$ gains some degeneracies for $j>1/2$.
For example, $|+1,-1\rangle+|-1,+1\rangle$ and $|0,0\rangle$ are both symmetric states that are also eigenstates of $J_z$ with eigenvalue $m_z=0$.
Since $J_z$ is nondegenerate within any $\SU(2)$ irrep, this means the symmetric subspace of 2 spin-1 systems must decompose into multiple $\SU(2)$ irreps.

\begin{figure}
    \centering
     \includegraphics[width=\columnwidth]{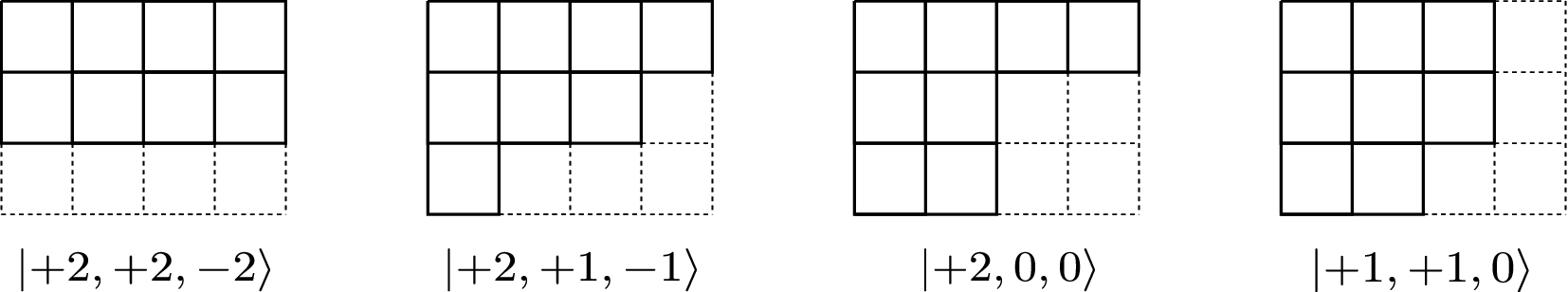}
    \caption{Restricted Young diagram showing a basis for the three-dimensional subspace of the totally symmetric subspace of 3 spin-$2$ systems for which $J_z=2$.
    The associated states are obtained by converting the number of boxes in each row to a $J_z$ eigenvalue by subtracting $j=-2$.
    Once symmetrized over the three subsystems, these states form a basis for the $J_z=2$ symmetric subspace.}
    \label{fig:restricted-young}
\end{figure}

A useful perspective on the decomposition is to consider the symmetric subspace as $n$ bosonic modes with at most $2j$ bosons in each mode~\cite{silva2022permutational}.
Each mode is associated with one of the spins, and the number of bosons in a mode corresponds to the $J_z$ eigenvalue of the associated spin (adding $j$ to the eigenvalue so the number of bosons ranges from $0$ to $2j$).
The total $J_z$ eigenvalue is then given by the total number of bosons, and the degeneracy of that eigenvalue in the symmetric subspace is given by the number of partitions of those bosons into $n$ distinct modes, restricted to putting no more than $2j$ bosons in a single mode.
These can be counted using restricted Young diagrams, where the number of columns must not exceed $2j$ and the number of rows must not exceed $n$.
An example of such restricted Young diagrams and their associated states is given in \cref{fig:restricted-young}.

For example consider the symmetric subspace of $2$ spin-$1/2$ particles, where the symmetric subspace is spanned by the triplet states and has a total spin $J=1$ (the largest possible angular momentum under the tensor product).
Mapping this to the $2$ bosonic modes with at most $2j=1$ boson each, we enumerate all partitions of $N$ bosons among these modes for $N\in\{0,1,2\}$.
The possible partitions are given in \cref{tab:symmetric_subspace_qubit_1}.
Each total photon number $N$ corresponds only to a single restricted partition, consistent with our previous statement that the symmetric subspace is a single $\SU(2)$ irrep.

\begin{table}
    \centering
    \begin{tabular}{ |c| c c| }
    \hline
    $N$ & $n_1$ & $n_2$\\
    \hline
    0&   $0$ & $0$  \\ 
\hline
 1 &$1$ & $0$ \\ 
 \hline
 2&$1$ & $1$ \\  
 \hline
\end{tabular}
    \caption{The symmetric subspace of two spin $j=1/2$ for $n=2$ as two bosonic modes. $n_1$ and $n_2$ are the numbers of bosons in each of the modes (symmetrized combinations as they are bosons) and $N=n_1+n_2$.
    There is only one possible partition for each of the values of $N$ and accordingly, there exists only a single $\SU(2)$ irrep in the symmetric subspace.
    (Note that our restriction on the number of bosons allowed per mode disallows the partition of $2$ into $2,0$.)}
    \label{tab:symmetric_subspace_qubit_1}
\end{table}

\begin{table}
    \centering
    \begin{tabular}{ |c| c c|c c|  }
    \hline
    $N$ & $n_1$ & $n_2$ & $n_1$ & $n_2$\\
    \hline
    0&   $0$ & $0$ & &  \\ 
\hline
 1 &$1$ & $0$ & &  \\ 
 \hline
 2&$1$ & $1$ & 2 & 0\\  
 \hline
 3& $2$ & $1$& &  \\
 \hline 
 4& 2 & 2& &  \\
 \hline
\end{tabular}
    \caption{The symmetric subspace of $n=2$ spin $j=1$ systems.
    We find we need two columns to account for the distinct partitions of $N=2$ bosons.
    Filling in the columns from left to right for each $N$, we can identify the $\SU(2)$ irreps present by the number of occupied rows in each column.
    Here the first column has 5 occupied rows, corresponding to the 5-dimensional spin-2 irrep, and the second column has 1 occupied entry, corresponding to the spin-0 irrep.
    The particular partition of $N$ appearing in each column here has no special meaning, as the actual basis states of the irreps are generally superpositions of these partitions.}
    \label{tab:symmetric_subspace_qutrit_1}
\end{table}

As a first non-trivial example consider the case of spin $j=1$ and $n=2$.
The restricted partitions of bosons into two modes are given in \cref{tab:symmetric_subspace_qutrit_1}.
As we can see from the table there are two partitions of $N=2$ bosons into two modes, revealing a degeneracy of the $J_z$ operator for eigenvalue $m_z=0$.
Since a one-dimensional subspace of this degenerate subspace must belong to the spin-2 irrep, and there are no degeneracies for larger $m_z$, we see that the symmetric subspace decomposes into one copy of spin 2 and one copy of spin 0.

Using this same approach, one can numerically find that for the case of the tensor product of any two spin $j$, we could find that, 
\begin{equation}
j\otimes j  \overset{\mathrm{s.s}}{=} 2j\oplus (2j-2) \oplus (2j-4) \oplus \cdots.
\label{eq:symmetric_subspace_two_spins}
\end{equation}
Simple counting of the total dimensions verifies this and is given in detail in \cref{sec:symmetric_subspace_tensor_product}.

\begin{table}
\begin{center}
\begin{tabular}{| c| c c c |c c c|}
\hline
    $N$ & $n_1$ & $n_2$ & $n_3$ & $n_1$ & $n_2$ & $n_3$\\
    \hline
 0&   $0$ & $0$ & 0 & & &  \\ 
 \hline
 1 &$1$ & $0$& 0 & & & \\  
 \hline
 2 &$1$ & $1$& 0 & 2&0 &0\\
 \hline
 3 &$1$ & $1$& 1 &2 & 1 &0 \\
 \hline
 4& $2$ & $1$& 1& 2 & 2& 0 \\
 \hline
 5 & $2$ & $2$ &1 & & &\\
 \hline
 6& $2$ & $2$ &2& & &\\
 \hline
\end{tabular}
\end{center}
   \caption{The symmetric subspace of $n=3$ spin $j=1$.
   Three values of $N$ have multiple partitions, resulting in the second column having 3 occupied rows, and giving us a decomposition of the symmetric subspace into one copy of spin 3 and one copy of spin 1.}
    \label{tab:symmetric_subspace_j_1_n_3}
\end{table}

Similarly, we can use the same approach for more complex cases, for example, consider the case of $n=3$ and $j=1$, the possible restricted partitions are given in \cref{tab:symmetric_subspace_j_1_n_3}.
As we can see from the table we have two occupied columns with $d=7$ and $d=3$ which yields the two $\SU(2)$ irreps spin 3 and spin 1.

Since the specific symmetries we are interested in are only present for half-integer spins~\cite{gross_designing_2021}, the tensor product of two spins will not give us valid codespaces as it only produces integer spins.
Hence the first non-trivial cases of interest are three copies of a half-integer spin.
The decompositions into $\SU(2)$ irreps for the cases of $j=3/2,5/2,7/2, \text{ and } 9/2$ are given in \cref{eq:su(2)_three_spin_bosonic_picture}, where the bracket on top of the spins represent the multiplicity.

\begin{equation}
    \begin{aligned}
        \frac{3}{2}\otimes \frac{3}{2} \otimes \frac{3}{2} &\stackrel{\text{s.s}}{=} \frac{9}{2}\oplus \frac{5}{2} \oplus \frac{3}{2}\\
        \frac{5}{2}\otimes \frac{5}{2} \otimes \frac{5}{2} &\stackrel{\text{s.s}}{=} \frac{15}{2}\oplus \frac{11}{2} \oplus \frac{9}{2}\oplus \frac{7}{2}\oplus \frac{5}{2} \oplus \frac{3}{2}\\
          \frac{7}{2}\otimes \frac{7}{2} \otimes \frac{7}{2} &\stackrel{\text{s.s}}{=} \frac{21}{2}\oplus \frac{17}{2} \oplus \frac{15}{2}\oplus \frac{13}{2}\oplus \frac{11}{2}  \oplus \frac{9}{2}^{(2)}\\
          &\oplus \frac{7}{2}\oplus \frac{5}{2} \oplus \frac{3}{2}\\
          \frac{9}{2}\otimes \frac{9}{2} \otimes \frac{9}{2} &\stackrel{\text{s.s}}{=} \frac{27}{2}\oplus \frac{23}{2} \oplus \frac{21}{2}\oplus \frac{19}{2} \oplus \frac{17}{2}\oplus \frac{15}{2}^{(2)}\\
          &\oplus \frac{13}{2} \oplus \frac{11}{2}^{(2)}\oplus \frac{9}{2}^{(2)} \oplus \frac{7}{2} \oplus \frac{3}{2}\\
          \label{eq:su(2)_three_spin_bosonic_picture}
    \end{aligned}
\end{equation}

\section{Correcting small random $\mathrm{SU}(2)$ errors}
\label{sec:correcting_first_order_errors}

In the case of errors that are small random $\SU(2)$ rotations, the error operators to first order in the rotation angle will be linear in the angular-momentum operators $\{J_x, J_y, J_z\}$, or equivalently first rank tensor operators $T^{(1)}_q$ with $-1\leq q\leq 1$.
Tensor products of these errors to first order are permutations of
\begin{equation}
\mathcal{E}= A\otimes {1}\otimes {1}
\label{eq:error_tensor_product_structure}
\end{equation}
where $A \in \{J_x,J_y,J_z\}$ or $T^{(1)}_q$.
Thus the Knill-Laflamme conditions one needs to check are
\begin{equation}
\begin{aligned}
   & \bra{i}T^{1}_q\otimes T^{1}_{q'}\otimes \mathds{1}\ket{j}\\
   & \bra{i}T^{1}_q T^{1}_{q'}\otimes\mathds{1} \otimes \mathds{1}\ket{j}\\
   &\bra{i}T^{1}_q \otimes\mathds{1} \otimes \mathds{1}\ket{j},
   \label{eq:knill_Laflamme_conditons_first_order}
\end{aligned}
\end{equation}
where $i,j=\{0,1\}$ and $-1\leq q,q'\leq 1$.
However using the unitary operator,
\begin{equation}
    U_X^{\mathrm{tot}}=\bigotimes_{i}U_X
\end{equation}
where $U_X=\exp(-i\pi J_x)$ and for two states $\ket{\psi}$ and $\ket{\phi}$ which are real linear combinations of the states that respect the binary octahedral symmetry we get,
\begin{equation}
\bra{\psi}\otimes_{i}T^{k_i}_{-q_i}\ket{\phi}=(-1)^{\sum_i{q_i}}\bra{\phi}\otimes_{i}T^{k_i}_{q_i}\ket{\psi}
    \label{eq:property_22}
\end{equation}
which leaves us with  the error correction conditions
\begin{widetext}
\begin{equation}
\begin{aligned}
\bra{0}\otimes_{i}T^{k_i}_{q_i}\ket{0}=&(-1)^{\sum_i k_i-\sum_i q_i} \bra{1}\otimes_{i}T^{k_i}_{q_i}\ket{1}\implies \text{ only consider } \left(\sum_ik_i \in \text{ odd and }\sum_iq_i\equiv 0\text { mod } 4\right),\\
\bra{0}\otimes_{i}T^{k_i}_{q_i}\ket{1}=&(-1)^{\sum_i k_i-\sum_i q_i} \bra{0}\otimes_{i}T^{k_i}_{q_i}\ket{1}\implies \text{ only consider } \left(\sum_i k_i \in \text{ odd and }\sum_i q_i  \equiv 1\text { mod } 4\right).
\end{aligned}
\label{eq:knill_Laflamme_conditions_tensor}
\end{equation}
\end{widetext}
Where we used the fact that the tensor product of spherical tensors shifts the total angular momentum by the sum of the individual shifts,
\begin{equation}
\begin{aligned}
 &\otimes_{i} T^{k_i}_{q_i}\ket{j_z=m_1,j_z=m_2,\hdots,j_z=m_N}\\
 &\propto\ket{j_z=m_1+q_1,j_z=m_2+q_2,\hdots,j_z=m_N+q_N},
\end{aligned}
\end{equation}
and hence the spacing arguments we used to get the $\text{mod } 4 $ are still valid for a code respecting the binary octahedral group.

Turning our attention back to the case of the Knill-Laflamme conditions for the first-order errors in the angular momentum operators in \cref{eq:knill_Laflamme_conditons_first_order}, the condition  $\bra{i}T^{1}_q\otimes T^{1}_{q'}\otimes \mathds{1}\ket{j}$ is trivially satisfied when $\sum_i k_i$ is even.
Now using the fact that when one multiplies two spherical tensors of rank $k_1, k_2$ the decomposition consists of all the spherical tensors with rank $k$, where $\lvert k_1-k_2 \rvert\leq k \leq k_1+k_2$, the condition 
\begin{equation}
    \bra{i}T^{1}_q T^{1}_{q'}\otimes\mathds{1} \otimes \mathds{1}\ket{j}
\end{equation}
leaves us with spherical tensors with rank $0,1,2$.
However, from \cref{eq:knill_Laflamme_conditons_first_order} the rank $0$ and $2$ cases are trivially satisfied, and hence the only term to check is $\bra{i}T^{1}_q\otimes {1}\otimes{1}\ket{j}$.
We recall that, when correcting for total angular momentum errors on binary octahedral codes, it was sufficient to check
\begin{align}
    \bra{0}J_{z,\text{total}}\ket{0}
    &=
    0
    \,.
\end{align}
Since we're considering codes in the symmetric subspace, we have
\begin{align}
    \tfrac{1}{3}\bra{0}J_{z,\text{total}}\ket{0}
    &=
    \bra{0}J_z\otimes\mathds{1}\otimes \mathds{1}\ket{0}
    \\
    &=
    \bra{0}\mathds{1}\otimes J_z\otimes \mathds{1}\ket{0}
    \\
    &=
    \bra{0}\mathds{1}\otimes \mathds{1}\otimes J_z\ket{0}
\end{align}
so correcting first-order single-system angular momentum errors in a binary octahedral code is equivalent to correcting first-order global angular-momentum errors.

\subsection{Case of three $j=3/2$}
According to \cref{eq:su(2)_three_spin_bosonic_picture} the symmetric subspace of three spin-$3/2$ systems decomposes into three $\SU(2)$ irreps.
Faithful two-dimensional binary-octahedral irreps are present both in the $j=9/2$ and the $j=5/2$ $\SU(2)$ irreps.
However, these irreps are incompatible with each other.
In the notation of~\cite{gross_designing_2021}, $j=9/2$ has a single copy of $\varrho_4$ while $j=5/2$ has a single copy of $\varrho_5$.
While this prevents us from engineering a code with binary-octahedral symmetry, one obtains more freedom by relaxing to binary-tetrahedral symmetry~\cite{gross_designing_2021}.

For the binary tetrahedral symmetry, the error condition becomes,
\begin{widetext}
\begin{equation}
\begin{aligned}
\bra{0}\otimes_{i}T^{k_i}_{q_i}\ket{0}=&(-1)^{\sum_i k_i-\sum_i q_i} \bra{1}\otimes_{i}T^{k_i}_{q_i}\ket{1}\implies \text{ only consider } \left(\sum_ik_i \in \text{ odd and }\sum_iq_i\equiv 0\text { mod } 2\right),\\
\bra{0}\otimes_{i}T^{K_i}_{q_i}\ket{1}=&(-1)^{\sum_i k_i-\sum_i q_i} \bra{0}\otimes_{i}T^{k_i}_{q_i}\ket{1}\implies \text{ only consider } \left(\sum_i k_i \in \text{ odd and }\sum_i q_i  \equiv 1\text { mod } 2\right).
\end{aligned}
\label{eq:knill_Laflamme_conditions_tensor_tetrahedral}
\end{equation}
\end{widetext}
the factor of $\text{mod }2 $ appears as the spacing of the binary tetrahedral code words is $2$ instead of the $4$ for the binary octahedral codewords.
However, for the case of first-order errors in the angular momentum, the only non-trivial condition we need to satisfy is $\bra{i}T^{1}_q\otimes {1}\otimes{1}\ket{j}$.


Making this relaxation, we find that $j=9/2$ and $j=5/2$ each have a copy of the faithful two-dimensional binary-tetrahedral irrep $\varrho_4$ (again in the notation of the appendix of~\cite{gross_designing_2021}).
The expectation values of $J_z$ for the logical 0s of these two irreps have opposite signs, so we engineer a combined codeword with vanishing $J_z$ expectation value to satisfy the error-correction conditions:
\begin{equation}
    \ket{0}_{}=\frac{1}{\sqrt{16}}\left(\sqrt{5}\ket{0}_\frac{9}{2}+\sqrt{11}\ket{0}_\frac{5}{2}\right).
\end{equation}
where
\begin{equation}
    \begin{aligned}
        \ket{0}_\frac{9}{2}
        &=
        \frac{\sqrt{6}}{4}\ket{\frac{9}{2},\frac{9}{2}}
        +\frac{\sqrt{21}}{6}\ket{\frac{9}{2},\frac{1}{2}}
        +\frac{\sqrt{6}}{12}\ket{\frac{9}{2},\frac{-7}{2}},
        \\
        \ket{0}_\frac{5}{2}
        &=
        -\frac{\sqrt{6}}{6}\ket{\frac{5}{2},\frac{5}{2}}
        +\frac{\sqrt{30}}{6}\ket{\frac{5}{2},\frac{-3}{2}}.
    \end{aligned}
\end{equation}
The projectors onto the irreps in $j=9/2$ and $j=5/2$ can be constructed from the character for $\varrho_4$ along with the representatives for the binary-tetrahedral group elements provided by the $\SU(2)$ irreps as discussed in~\cite{gross_designing_2021}.

\subsection{Case of three $j=5/2$}

Next, consider the case of three spin $5/2$ whose symmetric-subspace decomposition is also given in \cref{eq:su(2)_three_spin_bosonic_picture}.
Again we are looking for multiple copies of one of the faithful two-dimensional irreps of the binary-octahedral group.
For this case, we have multiple options and for simplicity we chose the irrep $\varrho_4$ appearing in $j=9/2$ and $j=11/2$.
The corresponding logical zero states are
\begin{equation}
\begin{aligned}
\ket{0}_{\frac{11}{2}}&=\frac{\sqrt{21}}{12}\ket{\frac{11}{2};\frac{9}{2}} -\frac{\sqrt{2}}{4}\ket{\frac{11}{2};\frac{1}{2}}+\frac{\sqrt{105}}{12}\ket{\frac{11}{2};\frac{-7}{2}},\\
\ket{0}_{\frac{9}{2}}&=\frac{\sqrt{6}}{4}\ket{\frac{9}{2};\frac{9}{2}}+\frac{\sqrt{21}}{6}\ket{\frac{9}{2};\frac{1}{2}}+\frac{\sqrt{6}}{12}\ket{\frac{7}{2};\frac{1}{2}},
\end{aligned}
\end{equation}
These codewords have equal and opposite expectation values
\begin{equation}
\begin{aligned}
\bra{0} J_z\otimes\mathds{1}\otimes \mathds{1}\ket{0}_{\frac{11}{2}}=&-\frac{11}{18}\\
\bra{0} J_z\otimes\mathds{1}\otimes \mathds{1}\ket{0}_{\frac{9}{2}}=&\frac{11}{18}\\
\end{aligned}
\end{equation}
meaning we get a codeword that corrects for first-order errors by simply taking a uniform superposition:
\begin{equation}
\ket{0}_L=\frac{1}{\sqrt{2}}\left(\ket{0}_{\frac{11}{2}}+\ket{0}_{\frac{9}{2}}\right).
\end{equation}

\section{Correcting Optical Pumping}
\label{sec:correcting_second_order_errors}

In the case of the error that is similar to optical pumping~\cite{deutsch2010quantum}, the error operators are of the form $J_i^{l}J_j^{m}$, where $\{i,j=x,y,z\}$ and $l+m\leq 2$. 
However, we find it convenient again to express these errors in terms of the spherical tensors $\{T_q^{k};  -k\leq q\leq k\}$ as they form  an orthogonal basis for errors and can be written in terms of angular momentum operators as given in \cref{sec:spherical_tensors}.
Errors of this type acting on a single spin are permutations of 
\begin{equation}
\mathcal{E}= A\otimes \mathds{1}\otimes \mathds{1}
\end{equation}
where $A \in \{T_q^{k}; 1\leq k\leq 2,-k\leq q\leq k\}$ . 
We see the Knill-Laflamme conditions in \cref{eq:knill_Laflamme_conditions_tensor} are trivially satisfied except the ones given in \cref{tbl:errors_second_order}. 
The errors with total $\sum k \text{ mod } 2 =0$ are trivially satisfied by \cref{eq:knill_Laflamme_conditions_tensor}.

In our numerical simulations, we observed that we either need to satisfy the diagonal or off-diagonal condition for the codes respecting the binary octahedral symmetry. 
Thus if one finds a code satisfying the diagonal conditions the off-diagonal conditions will be trivially satisfied and vice-versa, which is also true  for the error operators that are linear in the angular-momentum operators. 
Unlike the case of linear angular-momentum errors finding the codeword analytically is hard and one needs to rely on numerical methods to find the codewords; the method is described in detail in \cref{sec:Algorithm}.
Also as one is interested in the local rather than global errors we need to transform the basis from the $\ket{j_{\mathrm{tot}},j_z^{\mathrm{tot}}} \to \ket{j_1,m_1;j_2,m_2; j_3,m_3}$
using the Clebsch-Gordan coefficients where $\{j_i,m_i\}$ refers to the angular momentum basis of the individual spins.

From, \cref{eq:su(2)_three_spin_bosonic_picture}, there are multiple $\SU(2)$ irreps within the symmetric subspace of the threefold tensor product of spin-$j$ systems.
Decomposing these further into binary octahedral irreps gives us high multiplicities for the two faithful two-dimensional irreps and therefore many degrees of freedom with which to satisfy the error-correction conditions. 
For example, consider the case of  spin $j=7/2$.
A possible codeword obtained numerically for the $\varrho_4$ irrep \cite{gross_designing_2021} is
\begin{equation}
\begin{aligned}
\ket{0}&\propto \sqrt{\frac{70}{849}}\ket{0}_\frac{21}{2}
+\sqrt{\frac{1}{4468}} \ket{0}_\frac{17}{2}^{1}
+\sqrt{\frac{338}{1251}}\ket{0}_\frac{17}{2}^{2}\\
+&\sqrt{\frac{112}{479}}\ket{0}_\frac{15}{2}
+\sqrt{\frac{515}{1246}} \ket{0}_\frac{13}{2}.
\end{aligned}
\end{equation}
where $\ket{0}_\frac{17}{2}^{1}$ and $\ket{0}_\frac{17}{2}^{2}$ are orthogonal choices for $|0\rangle$ within the multiplicity-two $\varrho_4$ irrep of the binary-octahedral representation derived from $j=17/2$, where the degeneracy is broken by diagonalizing $J_z$ in the subspace spanned by the logical $|0\rangle$s.

Similarly for the case of the $j=9/2$ we can use the $\SU(2)$ irreps given in \cref{eq:su(2)_three_spin_bosonic_picture} and can find a code numerically as
\begin{equation}
\begin{aligned}
\ket{0}&\propto  -\sqrt{\frac{2}{439}}\ket{0}_\frac{27}{2}^{1}
+\sqrt{\frac{55}{739}}  \ket{0}_\frac{27}{2}^{2}
-\sqrt{\frac{216}{349}} \ket{0}_\frac{23}{2}^{1} \\
+&\sqrt{\frac{133}{1090}} \ket{0}_\frac{23}{2}^{2}
-\sqrt{\frac{237}{1316}}\ket{0}_\frac{21}{2}
\end{aligned}
\end{equation}
where again we have used the $\varrho_4$ irrep and where superscripts in the codeword represent the multiplicities for $j=27/2$ and $j=23/2$ and degeneracy is broken by diagonalizing $J_z$ in the subspace spanned by the logical $|0\rangle$s.

\begin{table}
  \centering
  \begin{tabular}{  r |  r }
  \hline
     diagonal errors & off-diagonal errors 
    \\
    \hline
    $ \bra{0} T_{0}^{1}\otimes\mathds{1}\otimes \mathds{1}\ket{0}_L$&$ \bra{0} T_{1}^{1}\otimes\mathds{1}\otimes \mathds{1}\ket{1}_L$ \\\\

$ \bra{0} T_{0}^{2}T_{0}^{1}\otimes\mathds{1}\otimes \mathds{1}\ket{0}_L$ & $\bra{0}T^1_{-1}\otimes T^2_{2}\otimes \mathds{1}\ket{1}_L$\\\\

$\bra{0}T^1_{-1}\otimes T^2_{1}\otimes \mathds{1}\ket{0}_L$ &$\bra{0}T^1_{1}\otimes T^2_{0}\otimes \mathds{1}\ket{1}_L$\\\\

$\bra{0}T^1_{1}\otimes T^2_{-1}\otimes \mathds{1}\ket{0}_L$&$\bra{0}T^1_{0}\otimes T^2_{1}\otimes \mathds{1}\ket{1}_L$\\\\

$\bra{0}T^1_{0}\otimes T^2_{0}\otimes \mathds{1}\ket{0}_L$&$\bra{0}T^1_{-1}\otimes T^2_{-2}\otimes \mathds{1}\ket{1}_L$\\\\

& $\bra{0}T^1_{-1}T^2_{-2}\otimes \mathds{1}\otimes \mathds{1}\ket{1}_L$\\\\
\hline
  \end{tabular}
  \caption{The relevant errors we need to satisfy for the error correction up to the second order for the tensor product of three spins. The table is constructed using the \cref{eq:knill_Laflamme_conditions} and the tensor product structure.
 }
  \label{tbl:errors_second_order}
\end{table}

Thus using the tensor-product structure of a minimum of $3$ spins with individual spins $j>1/2$ one can encode a qubit correcting the most significant error in these physical platforms, which are rotation errors and optical pumping.
This, in turn, provides an alternate approach for error correction with very low overhead, the number of physical systems to encode a logical qubit, by caring about the most significant error mechanisms.

\section{Correcting multibody errors with spin $j=\frac{1}{2}$}
\label{sec:symmetric_subspace_1_2}

Now we turn our attention to the case of the $N$-fold tensor product of $j=1/2$.
Here the only irrep in the symmetric subspace is spin $N/2$.
Hence we are shifting away from the paradigm of local (one-body) first- and second-order angular momentum errors, and will be considering non-local (multi-body) errors in this section. 
For this case, we can work with collective spin operators,
\begin{equation}
    J_{k}=\frac{1}{2}\sum_{i=1}^{N}\sigma_{k,i},
    \label{eq:collective_spin_operators}
\end{equation}
where $\sigma_{k,i}$ is the Pauli matrix acting on the $i$-th location and $k\in\{x,y,z\}$.

\noindent Using the property of the symmetric subspace in \cref{eq:permuatation_symmetry} we get 
\begin{equation}
    \langle J_k\rangle=\frac{N}{2} \langle \sigma_{k,1}\rangle=\frac{N}{2} \langle \sigma_{k,2}\rangle=\hdots=\frac{N}{2} \langle \sigma_{k,N}\rangle.
\end{equation}

Thus making the expectation value of the collective spin operator vanish makes all the local expectation values vanish which is the condition we studied for small random $\SU(2)$ errors in \cref{sec:correcting_first_order_errors}.

\noindent Now looking for codes for the qubit with the capacity to correct individual qubit errors, one can think of the same in terms of the collective spin operators. 
For example consider the case of the code corrects for all single body Pauli errors i.e. a code with distance $3$, the Knill-Laflamme conditions one needs to consider 
\begin{equation}
    \begin{aligned}
    &\bra{i} \sigma_{k,p} \ket{j}\\
    &\bra{i} \sigma_{k,p} \sigma_{l,p'} \ket{j},
    \end{aligned}
\end{equation}
where we used the fact that $\left(\sigma_{k,i}\right)^2=\mathds{1}$ and $p,p'=\{1,2,\hdots, N\}$, $k,l=\{x,y,z\}$.
However, if we restrict ourselves to the case of the codes respecting the binary octahedral symmetry and using the error correction condition derived in \cref{eq:knill_Laflamme_conditions_tensor} where we have all the operators with rank $k_i=1$, the only conditions remaining to check are
\begin{equation}
    \bra{i} \sigma_{k,p} \ket{j}_=\frac{2}{N}\bra{i} J_k \ket{j}.
\end{equation}

\noindent However for the binary octahedral symmetry for the collective spin operators the only condition we need to satisfy is \cite{gross_designing_2021},
\begin{equation}
    \bra{0}_{}J_z\ket{0}_{}.
    \label{eq:first_order_error_correction}
\end{equation}
For example one can think of a code with parameter $\left[\left[n,k,d\right]\right]=\left[\left[13,1,3\right]\right]$ in the $\varrho_5$ irrep for the octahedral symmetry and the codeword is,
\begin{equation}
    \ket{0}_=\frac{\sqrt{105}}{14}\ket{0}_0+\frac{\sqrt{91}}{14}\ket{0}_{1},
\end{equation}
where the states in the basis $\ket{J,J_z}$ is,
\begin{widetext}
\begin{equation}
    \begin{aligned}
        \ket{0}_0={}&\frac{\sqrt{910}}{56}\ket{\frac{13}{2},\frac{13}{2}}-\frac{3\sqrt{154}}{56}\ket{\frac{13}{2},\frac{5}{2}}       -\frac{\sqrt{770}}{56}\ket{\frac{13}{2},-\frac{3}{2}}+\frac{\sqrt{70}}{56}\ket{\frac{13}{2},-\frac{11}{2}}\\
         \ket{0}_1={}&\frac{\sqrt{231}}{84}\ket{\frac{13}{2},\frac{13}{2}}-\frac{3\sqrt{1365}}{84}\ket{\frac{13}{2},\frac{5}{2}}
       -\frac{\sqrt{273}}{84}\ket{\frac{13}{2},-\frac{3}{2}}+\frac{\sqrt{3003}}{84}\ket{\frac{13}{2},-\frac{11}{2}}.
    \end{aligned}
\end{equation}
\end{widetext}

Next, we can consider the case of the error correcting code that corrects two Pauli errors, otherwise known as a distance-$5$ code. 
We start by considering correcting global angular-momentum errors up to the second order.
The octahedral symmetry of the codes reduces the Knill-Laflamme conditions \cref{eq:knill_Laflamme_conditions} we need to satisfy to
    \begin{align}
    \bra{i}J_z\ket{j}&=C_z\delta_{ij},
    \label{eq:first_condtion} \\
     \bra{i} J_z^3\ket{j}&=C_{zz}\delta_{ij}
     \label{eq:second_condtion},\\
     \bra{i} J_zJ_x^2\ket{j}&=C_{xz}\delta_{ij},
      \label{eq:third_condtion}\\
     \bra{i} J_xJ_yJ_z\ket{j}&=C_{xyz}\delta_{ij}
     \label{eq:fourth_condtion},
    \end{align}
where $i,j=\{0,1\}$.
Now as we have seen in \cref{sec:introduction_to_binary_octahedral_code} the condition $\bra{i} J_z\ket{j}_{}$ is equivalent to just satisfying $\bra{0} J_z\ket{0}=0$.
Again invoking the support structure of octahedral codes in \cref{sec:introduction_to_binary_octahedral_code} and the operator $U_X$ defined in \cref{eq:Ux_property} yields
\begin{equation}
    \begin{aligned}
    \bra{0} J_z^3\ket{1}&=\bra{1} J_z^3\ket{0}=0\\
    \bra{0}J_z^3\ket{0}&=-\bra{1} J_z^3\ket{1}.
    \end{aligned}
\end{equation}
Thus the condition need to satisfy the \cref{eq:second_condtion} reduces to  $\bra{0}J_z^3\ket{0}=0$.\\
\noindent Now using the fact that $J_\pm=J_x\pm i J_y$, we get
\begin{equation}
    J_x^2=\frac{1}{4}\left(J_+^2+J_-^2+2j(j+1)\mathds{1}+2J_z^2\right),
\end{equation}
therefore $
     J_zJ_x^2  =\frac{1}{4}\left(J_z J_+^2 + J_z J_-^2 +2j(j+1)\mathds{1} + J_z^3\right)$. 
Again invoking the support property of the binary octahedral symmetry yields
\begin{equation}
    \begin{aligned}
    \bra{0}J_z J_{\pm}^2\ket{1}&=\bra{1}J_z J_{\pm}^2\ket{0}=0\\
    \bra{0} J_z J_{\pm}^2\ket{0}&=\bra{1} J_z J_{\pm}^2\ket{1}=0.
    \end{aligned}
    \label{eq:another_condition}
\end{equation}
Thus to satisfy \cref{eq:third_condtion} it is sufficient to satisfy \cref{eq:second_condtion}.\\
\noindent Now for \cref{eq:fourth_condtion} one can use
\begin{equation}
    J_xJ_y=\frac{-i}{4}\left(J_+^2-J_-^2-J_z\right)
\end{equation}
to show $
  J_xJ_yJ_z=\frac{-i}{4}\left(J_+^2J_z-J_-^2 J_z-J_z^2\right)$.
However, from \cref{eq:another_condition}, and using
\begin{equation}
    \begin{aligned}
    \bra{0} J_z^2\ket{1}_{}&=\bra{1} J_z^2\ket{0}_{}=0\\
    \bra{0} J_z^2\ket{0}_{}&=\bra{1} J_z^2\ket{1}_{}
    \end{aligned}
\end{equation}
from \cite{gross_designing_2021}, we see that \cref{eq:fourth_condtion} is trivially satisfied, and thus to correct all the errors up to second power in angular momentum one only needs to satisfy
\begin{equation}
    \begin{aligned}
         \bra{0}J_z\ket{0}&=0\\
         \bra{0}J_z^3\ket{0}&=0.
    \label{eq:second_order_condition}
    \end{aligned}
\end{equation}
Armed with this result, we turn our attention to the local errors that actually concern us.
For a collection of spin $1/2$ systems,
\begin{equation}
    \begin{aligned}
    J_z^2=&\frac{1}{4}\sum_{i,j}\sigma_{z,i}\sigma_{z,j}\\
    =&\frac{1}{4}\sum_{i=j} \mathds{1}+\frac{1}{4}\sum_{i\neq j} \sigma_{z,i}\sigma_{z,j}.
    \end{aligned}
\end{equation}
Again using the fact that $\left(\sigma_{z,i}\right)^2=\mathds{1}$ we get
\begin{equation}
        \begin{aligned}
    J_z^3&=\frac{1}{8}\sum_{i,j,k}\sigma_{z,i}\sigma_{z,j}\sigma_{z,k}\\
    &=\frac{1}{8}\left(4\sum_k \sigma_{z,k}+\sum_{i\neq j\neq k} \sigma_{z,i}\sigma_{z,j}\sigma_{z,k}\right).
    \end{aligned}
\end{equation}
For a state in the symmetric subspace for $N$ spins,
\begin{equation}
    \begin{aligned}
    \langle J_z^3\rangle &= \frac{1}{8}\left(4\sum_k \langle \sigma_{z,k} \rangle+\sum_{i\neq j\neq k} \langle \sigma_{z,i}\sigma_{z,j}\sigma_{z,k}\rangle\right)\\
    &=\langle J_z\rangle+N(N-1)(N-2)\langle \sigma_{z,1}\sigma_{z,2}\sigma_{z,3}\rangle
    \end{aligned}
\end{equation}
Thus if we have a code that satisfies \cref{eq:second_order_condition}, the code the Knill-Laflamme conditions of the form $\sigma_{z,i}\sigma_{z,j}\sigma_{z,k}$. 
Now consider a general Knill-Laflamme condition, 
\begin{equation}
    \bra{i}_{}\sigma_{p,k}\sigma_{q,l}\sigma_{r,m}\ket{j}_{},
    \label{eq:condtion_qubit_1}
\end{equation}
where $p,q,r=\{x,y,z\}$ and $k,l,m=\{1,2,\hdots, N\}$ for $N$ spin $1/2$ systems.
One can again look at the collective spin operators and the expansion of $J_xJ_yJ_z$ and $J_zJ_x^2$ in terms of Pauli operators.
We have,
\begin{equation}
    J_xJ_yJ_z=\frac{1}{8}\sum_{i,j,k}\sigma_{x,i}\sigma_{y,j}\sigma_{z,k}.
\end{equation}
now using the fact that $\sigma_x=\sigma_{+}+\sigma_{-}$ and $\sigma_{y}=-i \left(\sigma_+-\sigma_{-}\right)$,
\begin{equation}
\begin{aligned}
 J_xJ_yJ_z&=\frac{-i}{8}\left(\sum_{i,j,k}\sigma_{+,i}\sigma_{+,j}\sigma_{z,k}-\sigma_{-,i}\sigma_{-,j}\sigma_{z,k}\right)\\
 &+\frac{i}{8}\left(\sum_{i,j,k}\sigma_{+,i}\sigma_{-,j}\sigma_{z,k}-\sigma_{-,i}\sigma_{+,j}\sigma_{z,k}\right).
\end{aligned}
   \end{equation}
   However, the Knill-Laflamme condition for the first two terms is trivially satisfied using \cref{eq:knill_Laflamme_conditions_tensor} and we need not consider the case when either $i,j,k$ are repeated as the total rank $\sum_i k_i$ is even for that case and those cases are trivially satisfied again by \cref{eq:knill_Laflamme_conditions_tensor}.
   Thus the only non-trivial terms to consider are,
   \begin{equation}
   \begin{aligned}
   &\frac{i}{8}\left(\sum_{i,j,k}\sigma_{+,i}\sigma_{-,j}\sigma_{z,k}-\sigma_{-,i}\sigma_{+,j}\sigma_{z,k}\right)\\
   &=i\comm{J_+}{J_{-}}J_z\\
   &=-J_z^2.
   \end{aligned}
   \end{equation}
   Thus the condition for $\sigma_{x,i}\sigma_{y,j}\sigma_{z,k}$ is satisfied if the global condition for $J_z^2$ is satisfied and for the binary octahedral symmetry the condition for $J_z^2$ is trivially satisfied.
   Now we can look at the expansion of $J_zJ_x^2$ and we get,
\begin{equation}
    \begin{aligned}
    J_zJ_x^2=\frac{1}{8} \sum_{i,j,k}\sigma_{z,i}\sigma_{x,j}\sigma_{x,k},
    \end{aligned}
\end{equation}
again expanding the $\sigma_{x}$ and ignoring the trivially satisfied cases we are left with the terms,
\begin{equation}
    \begin{aligned}
     &\frac{1}{8}\left(\sum_{i,j,k}\sigma_{+,i}\sigma_{-,j}\sigma_{z,k}+\sigma_{-,i}\sigma_{+,j}\sigma_{z,k}\right)\\
     &=2J_z (J_x^2+J_y^2)\\
     &=2J_z^3+2J_z\left(j(j+1)\right)
    \end{aligned}
\end{equation}
where $j=N/2$, is the spin of the totally symmetric subspace.
Thus if we satisfy the global condition of $J_z^3$ and $J_z$, the condition for $\sigma_{z,i}\sigma_{x,j}\sigma_{x,k}$ is satisfied, and hence only condition we need to check to satisfy all the errors up to distance $5$ is to check the global conditions given in \cref{eq:second_order_condition}.


The minimum spin we need to find conditions to correct for $J_z$ and $J_z^3$ is $j=25/2$  in the $\varrho_4$ irrep, 
i.e we need $25$ qubits and form a $\left[\left[25,1,5\right]\right]$ code. 
The codeword is approximately
\begin{equation}
    \ket{0} \propto   -\sqrt{\frac{267}{1213}}  \ket{0}_1+\sqrt{\frac{701}{1457}}\ket{0}_2+\sqrt{\frac{337}{1128}}\ket{0}_3,
    \label{eq:code_octahedral_25_2}
\end{equation}
where
\begin{widetext}
\begin{small}
\begin{equation}
  \begin{aligned}
    \ket{0}_1&=    -\sqrt{\frac{1377}{4132}}\ket{\frac{25}{2} \frac{25}{2}}  -\sqrt{\frac{1}{674}}\ket{\frac{25}{2} \frac{17}{2}}  -\sqrt{\frac{109}{1169}}\ket{\frac{25}{2},\frac{9}{2}}-\sqrt{\frac{803}{1918}}\ket{\frac{25}{2},\frac{1}{2}} \\
    &    -\sqrt{\frac{103}{690}}\ket{\frac{25}{2},\frac{-7}{2}}  -\sqrt{\frac{1}{263}}\ket{\frac{25}{2},-\frac{13}{2}}     -\sqrt{\frac{1}{3608}}\ket{\frac{25}{2},-\frac{21}{2}},\\
        \ket{0}_2&=   \sqrt{\frac{1}{4402}}\ket{\frac{25}{2} \frac{25}{2}}  -\sqrt{\frac{2}{839}}\ket{\frac{25}{2} \frac{17}{2}}  -\sqrt{\frac{293}{983}}\ket{\frac{25}{2},\frac{9}{2}}-\sqrt{\frac{11}{1264}}\ket{\frac{25}{2},\frac{1}{2}} \\
    &    \sqrt{\frac{913}{2925}}\ket{\frac{25}{2},\frac{-7}{2}}+  \sqrt{\frac{21}{412}}\ket{\frac{25}{2},-\frac{13}{2}}     -\sqrt{\frac{1069}{3264}}\ket{\frac{25}{2},-\frac{21}{2}},\\
     \ket{0}_3&=   -\sqrt{\frac{1}{61408}}\ket{\frac{25}{2} \frac{25}{2}}  +\sqrt{\frac{1750}{2781}}\ket{\frac{25}{2} \frac{17}{2}}  -\sqrt{\frac{325}{3548}}\ket{\frac{25}{2},\frac{9}{2}}
     +\sqrt{\frac{43}{763}}\ket{\frac{25}{2},\frac{1}{2}} \\
    &    -\sqrt{\frac{47}{551}}\ket{\frac{25}{2},\frac{-7}{2}}  +\sqrt{\frac{183}{1349}}\ket{\frac{25}{2},-\frac{13}{2}}     +\sqrt{\frac{2}{1011}}\ket{\frac{25}{2},-\frac{21}{2}}.\\
\end{aligned}  
\end{equation}
\end{small}
\end{widetext}

The distance $5$ code for the binary octahedral code has the same code parameters as the distance-$5$ surface code \cite{fowler2012surface,acharya2022suppressing}.
These codes have another interesting correspondence, in that they both belong to efficiently representable subsets of the full Hilbert space.
The codes we study in this article all belong to the symmetric subspace, which is spanned by the Dicke basis and has dimension $N+1$ which is linear instead of exponential in the number of qubits $N$.
The codewords for the surface code are stabilizer states, which we can efficiently represent by specifying a generating set of stabilizers of size $N-1$ \cite{gottesman1997stabilizer}.
One notable difference is that, unlike the surface code, the binary octahedral codes have full transversal single-qubit Cliffords.

\begin{figure}
    \centering
    \includegraphics[width=0.48\textwidth]{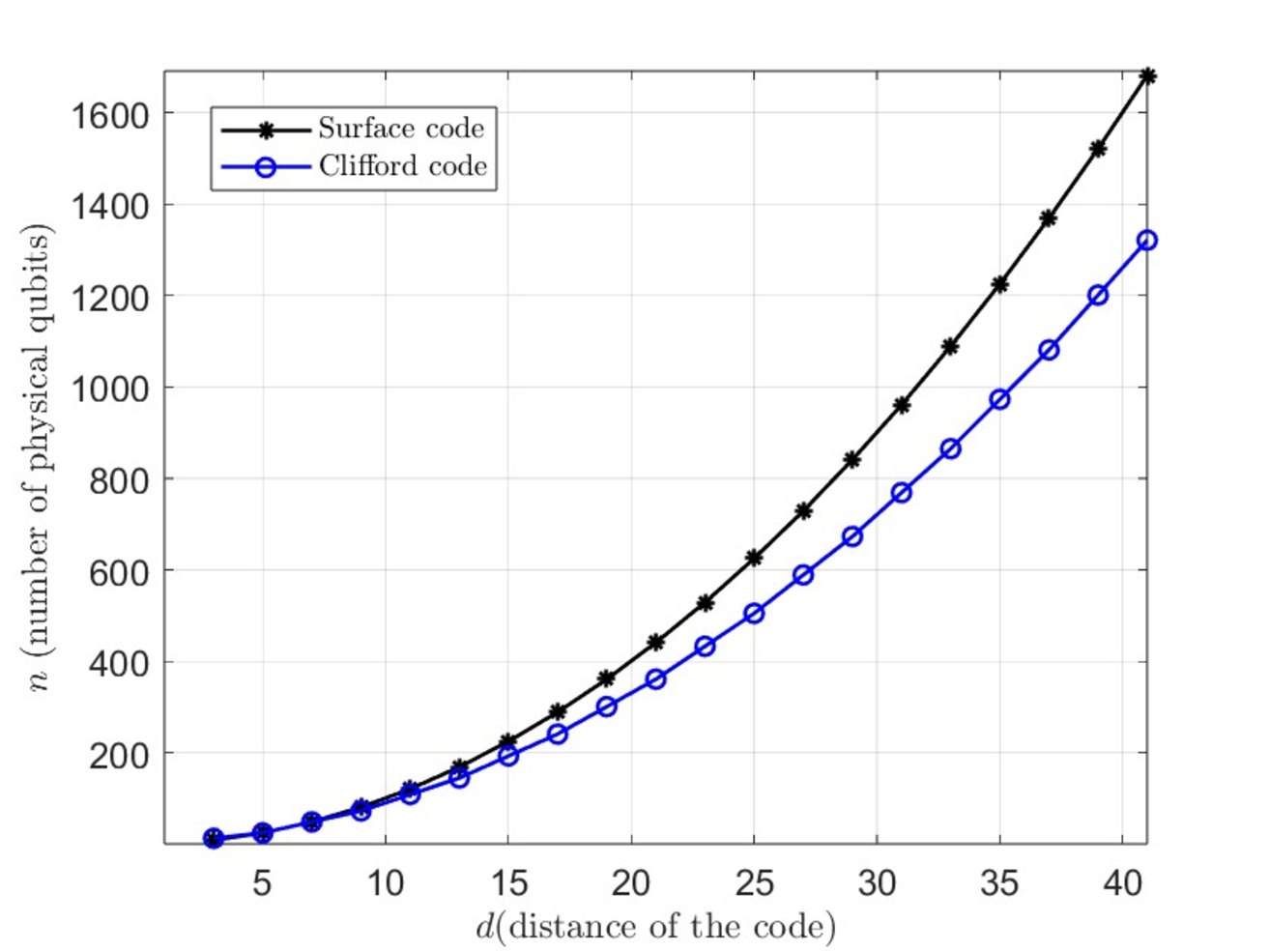}
    \caption{\textbf{Scaling of distance for Binary octahedral codes.} 
    The figure gives the number of physical qubits required for correcting errors up to a distance $d$ for the Surface code (rotated) and the binary octahedral code.}
    \label{fig:scalling_distance}
\end{figure}

Using the same approach as we did for the distance $3$ and $5$ codes one can build codes that have higher distances.
In \cref{fig:scalling_distance} the number of physical qubits as a function of distance is given for both the binary octahedral (Clifford) codes and the surface codes. 
Both scale quadratically in the distance, though the Clifford codes have an improved constant factor.

One can use the binary tetrahedral symmetry to find code words with even fewer qubits.
For example one can construct a $\left[\left[7,1,3\right]\right]$ code with codeword
\begin{equation}
    \ket{0}=\sqrt{\frac{{7}}{16}}\ket{0}_0+\sqrt{\frac{{16}}{16}}\ket{0}_{1},
\end{equation}
where
\begin{equation}
    \begin{aligned}
        \ket{0}_0&=-\frac{\sqrt{3}}{2}\ket{\frac{7}{2},\frac{5}{2}}+\frac{1}{2}\ket{\frac{7}{2},-\frac{3}{2}},\\
        \ket{0}_1&=\sqrt{\frac{{7}}{12}}\ket{\frac{7}{2},\frac{1}{2}}+\sqrt{\frac{5}{12}}\ket{\frac{7}{2},-\frac{7}{2}}.\\
    \end{aligned}
\end{equation}

The smallest distance $3$ stabilizer code that has transversal Cliffords is the Steane code \cite{steane1996multiple} with code parameters $[[7,1,3]]$
and also with binary octahedral symmetry, as it has transversal Clifford operators. 
The Steane code lies outside our classification as it does not live entirely within the symmetric subspace (being a superposition of spin $1/2$ and spin $7/2$), suggesting that more interesting codes might be found by looking beyond the symmetric subspace.

\section{conclusion and Outlook}
\label{sec:conlusions_and_future_work}

In this work, we focused on using binary octahedral symmetry to construct useful quantum error-correcting codes extending the ideas in \cite{gross_designing_2021}. 
In \cite{gross_designing_2021}, the codes were designed to protect against $\SU(2)$ errors in a single large spin. 
In this article, we developed a technique for designing codes for multiple copies of spins.
We leveraged the multiple $\SU(2)$ irreps within the symmetric subspace of the tensor product of several large spins to correct for the additional physically relevant error channel of tensor light shifts.
This resulted in numerically derived codes correcting tensor light shifts in three copies of spin $j=7/2$ and in three copies of spin $j=9/2$.

We derived general simplified error-correction conditions for correcting errors at arbitrary order using the structure of spherical tensors \cref{eq:knill_Laflamme_conditions,eq:knill_Laflamme_conditions_tensor}, which are polynomials of the angular momentum operators and well studied in the spin systems.

We additionally studied the case of qubits ($j=1/2$) and extended the framework to multi-body errors.
Again we used the symmetric subspace for a large number of spin $1/2$ systems and used the symmetries to find codes with distance $3$ for $n=7$ and distance $5$ for $n=25$.
The distance-$5$ code contrasts interestingly with the distance-$5$ surface code, which has the same code parameters but gives up the transversal Cliffords of the binary octahedral code in favor of its stabilizer structure.

The techniques outlined in this work can easily be extended to further develop codes with higher distances with octahedral symmetry.
An important open question is whether one can develop fault-tolerant schemes for these kinds of codes, as their highly non Abelian nature makes applying existing fault-tolerant strategies difficult.
Finally, it would be interesting to explore whether binary octahedral codes might have use as non stabilizer versions of the metrological codes discussed in~\cite{Faist21}.

\begin{acknowledgements}
The authors would like to acknowledge fruitful discussions with Ivan Deutsch and Milad Marvian about quantum error correction for spin systems. 
S.O. would like to acknowledge useful discussions with Tyler Thurtel during various stages of the work.
S.O. also thanks Pablo Poggi and Karthik Chinni for useful discussions about spherical tensors and their interesting properties, in particular Karthik for helping numerically create the spherical tensors as a polynomial of angular momentum operators.
The derivation of the counting of parameters for finding the symmetric subspace of two spins in the appendix was proven as a follow-up to a discussion with Austin Daniel.
This material is based upon work supported by the U.S. Department of Energy, Office of Science, National Quantum Information Science Research Centers, Quantum Systems Accelerator (QSA).
\end{acknowledgements}

\appendix
\section{Spherical Tensors}
\label{sec:spherical_tensors}
 The spherical tensor operators for a spin $j$ is defined in terms of the  commutator relations \cite{sakurai2014modern,klimov2008generalized},
\begin{equation}
    \begin{aligned}
      \comm{J_z}{T^{k}_q}&=q  T^{k}_q\\
      \comm{J_{\pm}}{T^{k}_q}&= \sqrt{k(k+1)-q(q\pm 1)} T^{k}_{q\pm 1}
    \end{aligned}
\end{equation}

\noindent Using the above relations  the irreducible spherical tensors can be explicitly written in terms of the angular momentum basis as \cite{sakurai2014modern,chinni2022reliability},
\begin{equation}
    T^{k}_q(j)=\sqrt{\frac{2k+1}{2j+1}}\sum_m \braket{j,m+q}{k,q;j,m} \ketbra{j,m+q}{j,m},
\end{equation}
where  $0\leq k \leq 2j$ and $-k\leq q \leq k$. 
The spherical tensor operators of rank $k$ can be expressed as order-$k$ polynomials in the angular-momentum operators \cite{chinni2022reliability,omanakuttan2022scrambling}.
The spherical tensor operators also form an orthonormal basis for the operators on an $\SU(2)$ irrep with respect to the Hilbert-Schmidt inner product:
\begin{equation}
    \Tr(T^{k_1}_{q_1}T^{k_2}_{q_2})=\delta_{k_1,k_2}\delta_{q_1,q_2}.
\end{equation}

Now  consider the unitary transformation given as,  $U_{X}=\exp(-i\pi J_x)$  which can also be written in terms of the angular momentum basis as,
\begin{equation}
    U_{X}=-i\sum_{m=-j}^{j}\ketbra{j,m}{j,-m}
\end{equation}
Thus the action of the unitary operator on the irreducible spherical tensor gives,
\begin{equation}
\begin{aligned}
U_{X} T_{q}^{k} U_{X}^{\dagger}=&\sum_{m=-j}^{j} \braket{j,m+q}{k,q;j,m} \ketbra{j,-m-q}{-m}
\end{aligned}
\end{equation}
Now using the transformation $m \to -m$ and using the fact that 
\begin{equation}
\braket{j,m+q}{k,q;j,m}=(-1)^{k}\braket{j,-m-q}{k,-q;j,-m},
\end{equation}
we get 
\begin{equation}
\begin{aligned}
U_{X} T_{q}^{k} U_{X}^{\dagger}=&(-1)^{k}\sum_m \braket{j,m-q}{k,-q;j,m} \ketbra{j,m-q}{j,m}\\=&(-1)^{k}T^{k}_{-q}.
\end{aligned}
\end{equation}
Thus the action of the $U_X$ on the spherical tensor operators is to flip the sign of $q$ and to add a rank-dependent phase of $\pm 1$ to the operator.

\section{Error correction condition}
\label{sec:real_code_word}
The logical Pauli Z operator on an irrep $\varrho$ of the binary octahedral group is given by~\cite{gross_designing_2021},
\begin{equation}
    \label{eq:logical-z}
    \sigma_z^{}=P_{\varrho} (i\exp(-i\pi J_z)) P_{\varrho}
\end{equation}
Logical $\ket{0}$ is taken to be a $+1$ eigenstate of the logical Pauli Z operator.
The projector for the binary octahedral group is given as
\begin{equation}
    P_{\varrho}=\frac{\mathrm{dim} \varrho}{|2 \mathrm{O}|}\sum_{g\in 2\mathrm{O}} \chi_\varrho(g)^{*}D(g).
\end{equation}
where $2\mathrm{O}$ is the single-qubit Clifford group \cite{gottesman1998heisenberg}, also called the binary octahedral group. 
Now from \cite{gross_designing_2021}, $\chi_\varrho(g)$ for the $\SU(2)$ irreps of interest are real.
For the binary octahedral group, we also have that every representative $D(g)$ is in the same conjugacy class as $D(g)^{\dagger}$, $D(g)^{T}$ and $D(g)^{*}$.
Restricting the sum to a fixed conjugacy class $[g]$ gives
\begin{align}
    \tfrac{1}{4}\chi_\varrho(g)^*\sum_{h\in[g]}
    \big(D(h)+D(h)^\dagger+D(h)^{T}+D(h)^*\big)
    \,.
\end{align}
The term for each conjugacy class is real-symmetric since $\chi_\rho$ is real and $D(g)+D(g)^\dagger+D(g)^T+D(g)^*$ is manifestly real and symmetric.
Thus we get  $P_{\varrho}$ to be a real symmetric matrix.
The term sandwiched by the projectors in \cref{eq:logical-z} is also real and symmetric for half-integer spins
\begin{equation}
    i\exp(-i\pi J_z)=(i\exp(-i\pi J_z))^{\dagger}=(i\exp(-i\pi J_z))^{T}
    \,,
\end{equation}
hence $\sigma_z^{}$ is a real-symmetric operator.

Now the eigenvector of a real symmetric matrix ($A$) can be found by solving the eigenvalue equation,
\begin{equation}
    (A-\lambda \mathds{1})\ket{\psi}=0
    \,.
\end{equation}
Since the eigenvalue $\lambda$ is real from $A$ being Hermitian, when solved by Gaussian elimination we get a real vector and hence the eigenvectors of a real symmetric matrix are also real (up to an overall constant which is not important).

Consider the following expectation value for two states $\ket{\psi}=\sum_{i}\alpha_i \ket{i}$ and $\ket{\phi}=\sum_i\beta_i\ket{i}$,  where $\ket{i}$ is in the angular momentum basis,
\begin{multline}
\bra{\psi}T_{-q}^{k}(j)\ket{\phi}=
\\
d_j^{k}\sum_{i,i^\prime,m} \alpha_i^{*}\beta_{i^\prime}C_{j,m,j,m-q}^{k,-q}
\braket{i^\prime}{j,m-q}\braket{j,m}{i}
\end{multline}
where $d_{j}^{k}=\sqrt{2k+1/2j+1}$ and
\begin{equation}
   C_{j_1,m_1 ,j_2,m_2}^{j_3,m_3}= \braket{j_3,m_3}{j_1,m_1;j_2,m_2} 
\end{equation}
is the Clebsch-Gordan coefficient. 
Now using the property that $ \braket{i}{j,m+q}=\braket{j,m+q}{i}$ as they are in both in the angular momentum basis.
\begin{multline}
     \bra{\psi}T_{-q}^{k}\ket{\phi}=
     \\d_j^{k}\sum_{i,i^{\prime},m} \alpha_i^{*}\beta_{i^\prime}C_{j,m,j,m-q}^{k,-q}  \braket{j,m-q}{i^{\prime}}\braket{i}{j,m}.
\end{multline}
   
Also, by transforming the above equation by $m\to m+q$ we get,
\begin{multline}
    \bra{\psi}T_{-q}^{k}\ket{\phi}=
    \\d_{j}^{k}\sum_{i,i^{\prime},m} \alpha_i^{*}\beta_{i^{\prime}}C_{j,m+q,j,m}^{k,-q}\braket{j,m}{i^{\prime}}\braket{i}{j,m+q}.
\end{multline}

Now using the property of the Clebsch-Gordan coefficients
\begin{equation}
  C_{j_1,m_1,j_2,m_2}^{j_3,m_3}  =(-1)^{j_1+j_2+j_3}C_{j_2,m_2,j_1,m_1}^{j_3,m_3},
\end{equation}
we get
\begin{multline}
     \bra{\psi}T_{-q}^{k}\ket{\phi}=\\
     (-1)^kd_j^{k}\sum_{i,i^\prime,m} \alpha_i^{*}\beta_{i^{\prime}}C_{j,m,j,m+q}^{k,-q} \braket{J,m}{i^\prime}\braket{i}{J,m+q}.
\end{multline}

Again using another property of Clebsch-Gordan coefficients,
\begin{equation}
 C_{j_1,m_1,j_2,m_2}^{j_3,m_3}=\sqrt{\frac{2j_1+1}{2j_2+1}}(-1)^{j_2+m_2}C_{j_1,m_1,j_2,m_2}^{j_3,-m_3}
\end{equation}
we get,
\begin{multline}
    \bra{\psi}T_{-q}^{k}\ket{\phi}=\\
    (-1)^qd_j^{k}\sum_{i,i^{\prime},m} \alpha_i^{*}\beta_{i^{\prime}}C_{j,m,j,m+q}^{k,q}\braket{j,m}{i^{\prime}}\braket{i}{j,m+q}.
\end{multline}

Since the computational-basis codewords for the binary octahedral case are real the amplitudes $\alpha_i$ and $\beta_i$ are real when $|\psi\rangle$ and $|\phi\rangle$ are computational-basis codewords, as when we're checking error-correction conditions, and thus
\begin{equation}
     \bra{\psi}T_{-q}^{k}\ket{\phi}=(-1)^q\bra{\phi}T_{q}^{k}\ket{\psi}.
\end{equation}

\section{Symmetric subspace under the tensor product of two spins} 
\label{sec:symmetric_subspace_tensor_product}
It is known that the $\SU(2)$ irrpes under the addition of two spin $j$ system is given as,
\begin{equation}
    j\otimes j=2j\oplus(2j-1)\oplus (2j-1)\oplus\cdots.
\end{equation}
Now focussing our attention on the symmetric subspace, in \cref{eq:symmetric_subspace_two_spins}, we numerically found that the symmetric subspace of two spin $j$ systems is composed of all $\SU(2)$ subspaces interleaving one in between starting from the highest possible angular momentum.
To verify this one could do dimension counting of these subspaces. 
First, consider the case of even multiple of spin $1/2$ and thus the dimension of the alternate $\SU(2)$ subspaces are given as, 
\begin{equation}
\begin{aligned}
\mathrm{dim}&=\sum_{k=0}^{j}4j+1-4k\\
&=4j(j+1)+j+1-2j(j+1)\\
&=2j^2+3j+1=\frac{(2j+1)(2j+2)}{2}\\
&=\mathrm{dim}\left(S_2(2j+1)\right).
\end{aligned}
\end{equation}
Now for the case of odd multiples of $1/2$ we have, 
\begin{equation}
\begin{aligned}
\mathrm{dim}&=\sum_{k=0}^{j-\frac{1}{2}}4j+1-4k\\
&=4j\left(j+\frac{1}{2}\right)+j+\frac{1}{2}-2\left((j-\frac{1}{2}\right)\left(j+\frac{1}{2}\right)\\
&=4j^2+2j+j+\frac{1}{2}-2j^2+\frac{1}{2}\\
&=2j^2+3j+1=\frac{(2j+1)(2j+2)}{2}\\
&=\mathrm{dim}\left(S_2(2j+1)\right),
\end{aligned}
\end{equation}
thus we get both for even and odd multiple of spin $1/2$ the dimension of the symmetric  subspace is $\SU(2)$ subspaces interleaving one in between starting from the highest possible angular momentum.

\section{Algorithm for finding the codeword for the case of second order errors}
\label{sec:Algorithm}
The simple algorithm for finding the codeword follows these three steps,\\
\textbf{Step I:}\\
Write the codewords as
\begin{equation}
\begin{aligned}
\ket{0}_L=\sum_{i=1}^n c_i \ket{0}_i,\, \ket{1}_L=\sum_{i=1}^n c_i \ket{1}_i,\
\end{aligned}
\end{equation}
where $i$ corresponds to the two-dimensional qubit spaces one has access to and $c_i\in \mathbf{R}$.\\
\textbf{Step II:}\\
Define the cost function,
\begin{equation}
\mathcal{F}[(\bm c)]=\sum_{\text{constraints}} |f(\bm c)|
\end{equation}
where $f(\bm c)$ is the value we get for each constraint we need to satisfy according to the Knill-Laflamme conditions in \cref{eq:knill_Laflamme_conditions}.\\
\textbf{Step II:}\\
Minimize the cost function to obtain the right codeword  where $\bm{c} \in \mathbf{R}^n$ such that,
\begin{equation}
  \bm{c}_{\mathrm{opt}}=   \underset{{\bm{c} \in \mathbf{R}}}{\text{arg min  }} \mathcal{F}[(\bm c)],
\end{equation}
which in turn gives the codewords as,
\begin{equation}
    \begin{aligned}
\ket{0}_L=\sum_i c_i^{\mathrm{opt}} \ket{0}_i, \, \ket{1}_L=\sum_i c_i^{\mathrm{opt}} \ket{1}_i.\
\end{aligned}
\end{equation}
\bibliography{references}

\end{document}